\documentclass[twocolumn,showpacs,preprintnumbers,amsmath,amssymb]{revtex4}

\usepackage{graphicx}
\usepackage{dcolumn}
\usepackage{bm}
\usepackage{amsmath}
\usepackage{amsthm}
\begin{document}
\title{Optimal discrimination between quantum operations}
\author{Lvzhou Li and Daowen Qiu}\email{issqdw@mail.sysu.edu.cn  (D.
Qiu).}
 \affiliation{%
 Department of
Computer Science, Zhongshan University, Guangzhou 510275,
 People's Republic of China
}%
\date{\today}

\begin{abstract}
In this paper, we address the problem of discriminating  two given
quantum operations. Firstly, based on the Bloch representation of
single qubit systems, we give the exact minimum error probability
of discriminating two single qubit quantum operations  by
unentangled  input states. In particular, for the Pauli channels
discussed in [Phys. Rev. A {\bf 71}, 062340 (2005)], we use a more
intuitional and visual method to deal with their discrimination
problem. Secondly, we consider the condition for perfect
discrimination of two quantum operations. Specially, we get that
two generalized Pauli channels are perfectly distinguishable if
and only if their characteristic vectors are orthogonal.
\end{abstract}
\maketitle
\section{Introduction}
Discrimination between quantum states is a fundamental task in
quantum information. It is well know that nonorthogonal quantum
states can not be discriminated perfectly. However, it is possible
to discriminate them in some relaxed ways. Typically, two
nontrivial discrimination schemes have been considered for
nonorthogonal states: one is the minimum-error  discrimination
\cite{Hel}, where each measurement outcome selects one of the
possible states and the error probability is minimum, and the
other is the optimal unambiguous discrimination \cite{Four}, where
unambiguity is paid by the possibility of getting inconclusive
results from the measurement. For a recent review on the problem
of discrimination between quantum states, we would like to refer
the reader to \cite{Bergou04}.

The problem of discrimination can also be applied to quantum
operations \cite{Nie}. In Refs.~\cite{Childs, Acin, Paris, Duan},
the authors considered discrimination between  unitary
transformations (special quantum operations), and they found that
entanglement-assisted input can not enhance the distinguishability
of unitary transformations. For the problem of discriminating
general quantum operations, however, there is not very much work,
and the first work on this problem may be owed to  Sacchi
\cite{MFS1, MFS2}. In Refs.~\cite{MFS1, MFS2}, the problem for
discriminating general quantum operations was firstly formulated,
and the problem for discriminating Pauli channels was also
addressed in detail. Specially, Sacchi \cite{MFS1, MFS2} showed
that unlike unitary transformations \cite{Childs, Acin, Paris},
entangled input states generally enhance the distinguishability of
general quantum operations. Also, Pauli channels, as a nontrivial
kind of quantum operations, were considered by \cite{MFS3} in the
approach of minimax discrimination. In addition, the unambiguous
scheme was also discussed for  general quantum operations by
\cite{Wang}, where a necessary and sufficient condition was given
for a given set of quantum operations to be unambiguously
distinguishable. By the way, recently there is another interesting
problem addressed by \cite{Chef07}, which considered unambiguous
discrimination among oracle operators.

In this paper, we focus on the problem of minimum-error
discrimination between two quantum operations ${\cal E}_1$ and
${\cal E}_2$.  This problem  can be reformulated into the problem
of finding in the input Hilbert space ${\cal H}$ the state $\rho$
such that the error probability in the discrimination of the
output states ${\cal E}_1(\rho)$ and ${\cal E}_2(\rho)$ is
minimum. We call such a strategy as {\it non-entanglement
strategy}.  Also, we have another strategy, called {\it
entanglement strategy}, where we can introduce
entanglement-assisted input states to increase the
distinguishability of the output states. In this case, the output
states to be discriminated will be of the form $({\cal
E}_1\otimes{\cal I})\rho$ and $({\cal E}_2\otimes{\cal I})\rho$,
where the input $\rho$ is generally a bipartite state of ${\cal
H}\otimes{\cal K}$, and the quantum operations act just on the
first party whereas the identity map ${\cal I}={\cal I}_{\cal K}$
acts on the second. It is suitable to  assume that ${\cal K}={\cal
H}$, and in the sequel, we always use this assumption.

We denote with $P_E$ the minimum error probability when a strategy
without ancilla is adopted. Then we have \cite{MFS2}
\begin{align}
P_E({\cal E}_1,{\cal E}_2)=\frac{1}{2}\Big(1-\max_{\rho\in{\cal
H}}||p_1{\cal E}_1(\rho)-p_2{\cal E}_2(\rho)||_1\Big),\label{P1}
\end{align}
where $p_1$ and $p_2$ are the {\it a priori} probabilities for
quantum operations ${\cal E}_1$ and ${\cal E}_2$, respectively,
and $||A||_1$ denotes the trace norm defined as
$||A||_1=\text{Tr}\sqrt{A^\dagger A}$. In particular, if $A$ is
Hermitian, then
\begin{align}
||A||_1=\sum_i|\lambda_i(A)|,
\end{align}
where $\lambda_i(A)$ denote the all eigenvalues of $A$. On the
other hand, by allowing the use of entangled input, we have
\cite{MFS2}
\begin{align}
P_E^{'}({\cal E}_1,{\cal
E}_2)=\frac{1}{2}\Big(1-\max_{\rho\in{\cal H}\otimes{\cal
K}}||p_1({\cal E}_1\otimes I)\rho-p_2({\cal E}_2\otimes
I)\rho||_1\Big).\label{P2}
\end{align}
 It is
readily seen that in both Eqs.~\eqref{P1} and \eqref{P2}, the
maximum value is achieved by  pure states, and thus, in the
following, we need only consider pure states as input states.

The remainder of this paper is organized as follows. In
Sec.~\ref{2}, firstly we give the exact minimum error probability
of discriminating two single qubit quantum operations with
non-entanglement strategy, and then by applying these results to
Pauli channels \cite{MFS1,MFS2}, we obtained a more intuitional
and visual solution for the problem of discriminating Pauli
channels. In Sec.~\ref{3}, we give a necessary and sufficient
condition for given two quantum operations to be perfectly
distinguishable. Furthermore, we get that two generalized Pauli
channels are perfectly distinguishable if and only if their
characteristic vectors are orthogonal. Finally, some conclusions
are made in Sec. ~\ref{4}.
\section{Discrimination of single qubit quantum operations
}\label{2} In the following, we address the problem of
minimum-error discrimination of single qubit quantum operations.
Indeed, we can make use of the Bloch representation \cite{Nie} of
single qubit systems to evaluate the value of $P_E$. Let ${\cal
H}_2$ denote the Hilbert space of a qubit system. Then it is well
know that any $\rho\in {\cal H}_2$ can always be written in the
form
\begin{align}
\rho=\frac{I+\Vec{r}\cdot\Vec{\sigma}}{2}
\end{align}
where $\Vec{r}=(r_x, r_y, r_z)$ is a three-dimensional real vector
with norm $||\Vec{r}||\leq 1$ ($||\cdot||$ denotes the {\it
Euclidean norm} on ${\bf C}^n$), and $\Vec{r}\cdot\Vec{\sigma}=
r_x\sigma_x+r_y\sigma_y+r_z\sigma_z$ with
$\{\sigma_x,\sigma_y,\sigma_z\}$ denoting the Pauli operators
\cite{Nie}. In this way, $\Vec{r}$  is called the {\it Bloch
vector} of $\rho$, and they have a one-to-one relation. Also, we
have that $\rho$ is a pure state if and only if $||\Vec{r}||=1$.

Based on the Bloch representation, we can visualize the effect of
any trace-preserving single qubit quantum operation ${\cal E}$
as the Bloch vector transformation \cite{Nie}
\begin{align}
\Vec{r}\rightarrow\Vec{r}\hskip 0.5mm'=M\Vec{r}+\Vec{c},
\end{align}
where $M$ is a $3\times 3$ real matrix, and $\Vec{c}$ is a
three-dimensional real vector, all of which can be computed by the
operator-sum representation of ${\cal E}$. Therefore, the quantum
operation ${\cal E}$ on  a single qubit is characterized by the
2-tuple $(M, \Vec{c})$. For the details, we refer to \cite{Nie}.

Next, we try to explore the minimum error probability for
discriminating two single qubit quantum operations. Before that,
we give some useful results below.

Firstly, we have that
\begin{align}
\frac{1}{2}(|a+b|+|a-b|)=\max\{|a|,|b|\}\label{max},
\end{align}
 where $a$ and $b$ are any real numbers.  This equation can be
easily verified  by discussing the two case: (1) $|a|\geq|b|$, (2)
$|a|<|b|$.

Secondly, we have a  useful lemma in the following.

 \noindent\textit {Lemma 1.}
 Let $\rho_1$ and $\rho_2$ be the
density operators of a single qubit system, with {\it a priori}
probabilities $p_1$ and $p_2$, respectively. Let $\Vec{r}_1$ and
$\Vec{r}_2$ be the Bloch vectors of  $\rho_1$ and $\rho_2$,
respectively. Then we have
\begin{align}
||p_1\rho_1-p_2\rho_2||_1=\max\Big\{|p_1-p_2|,||p_1\Vec{r}_1-p_2\Vec{r}_2||\Big\}.
\end{align}

\textit{ Proof.} First, we note that $\Vec{r}\cdot\Vec{\sigma}$
has eigenvalues $\pm||\Vec{r}||$. Then, by the Bloch
representation, we have
\begin{align}\begin{split}
&||p_1\rho_1-p_2\rho_2||_1\\
=&\frac{1}{2}||p_1(I+\Vec{r}_1\cdot\Vec{\sigma})-p_2(I+\Vec{r}_2\cdot\Vec{\sigma})||_1\\
=&\frac{1}{2}||(p_1-p_2)I+(p_1\Vec{r}_1-p_2\Vec{r}_2)\Vec{\sigma}||_1\\
=&\frac{1}{2}(|a-b|+|a+b|),
\end{split}\end{align}
where we let $a=p_1-p_2$, and $b=||p_1\Vec{r}_1-p_2\Vec{r}_2)||$.
Therefore, by Eq.~\eqref{max}, we have completed the proof.\qed

Now,  suppose that ${\cal E}_1$ and ${\cal E}_2$  are two single
qubit quantum operations, and  by the Bloch representation ${\cal
E}_1$ and ${\cal E}_2$ correspond to $(M_1,\Vec{c}_1)$ and
$(M_2,\Vec{c}_2)$, respectively. Then, the  minimum error
probability of discriminating ${\cal E}_1$ and ${\cal E}_2$ with
non-entanglement strategy can be evaluated as follows.
\begin{widetext}
\begin{align}
\begin{split}
P_E({\cal E}_1,{\cal
E}_2)=&\frac{1}{2}\Big(1-\max_{|\psi\rangle\in{\cal H}}||p_1{\cal
E}_1(|\psi\rangle\langle\psi|)-p_2{\cal E}_2(|\psi\rangle\langle\psi|)||_1\Big)\\
=&\frac{1}{2}\Big[1-\max_{||\Vec{r}||=1}\max\Big\{|p_1-p_2|,||p_1(M_1\Vec{r}+\Vec{c}_1)-p_2(M_2\Vec{r}+\Vec{c}_2)||\Big\}\Big]\\
=&\frac{1}{2}\Big[1-\max\Big\{|p_1-p_2|,
\max_{||\Vec{r}||=1}||(p_1M_1-p_2M_2)\Vec{r}+(p_1\Vec{c}_1-p_2\Vec{c}_2)||\Big\}\Big]\\
=&\frac{1}{2}\Big[1-\max\Big\{|p_1-p_2|,
\max_{||\Vec{r}||=1}||M\Vec{r}+\Vec{c}||\Big\}\Big], \label{Pe}
\end{split}
\end{align}
\end{widetext}
where we denote $(p_1M_1-p_2M_2)$ and
$(p_1\Vec{c}_1-p_2\Vec{c}_2)$ by $M$ and $\Vec{c}$, respectively.
Note that the  value $\max_{||\Vec{r}||=1}||M\Vec{r}+\Vec{c}||$
can be computed exactly for any fixed $M$ and $\Vec{c}$.
Therefore, the minimum error probability $P_E({\cal E}_1,{\cal
E}_2)$ given above can be evaluated for any two given quantum
operations on a single qubit system.

Below, we  show explicitly how to compute the value
$\max_{||\Vec{r}||=1}||M\Vec{r}+\Vec{c}||$ by  computing its
square. Without loss of generality, suppose that
\begin{eqnarray}
M=\left(%
\begin{array}{ccc}
  a_{11} &  a_{12} &  a_{13} \\
   a_{21} &  a_{22} &  a_{23} \\
   a_{31} &  a_{32} &  a_{33} \\
\end{array}%
\right),&\Vec{c}=\left(%
\begin{array}{c}
  c_1 \\
  c_2 \\
  c_3 \\
\end{array}%
\right),&\Vec{r}=\left(%
\begin{array}{c}
  x \\
  y \\
  z \\
\end{array}%
\right).
\end{eqnarray}
Then $||M\Vec{r}+\Vec{c}||^2=f(x,y,z)$, where function $f$ defined
as
\begin{align}
\begin{split}
f(x,y,z)=&(a_{11}x+a_{12}y+a_{13}z+c_1)^2\\
&+(a_{21}x+a_{22}y+a_{23}z+c_2)^2 \\
&+(a_{31}x+a_{32}y+a_{33}z+c_3)^2.
\end{split}
\end{align}
 Now, the  problem  reduces to finding out the maximal value of  $f(x,y,z)$
under the condition $x^2+y^2+z^2=1$. Clearly, this problem belongs
to the class of constrained extremum problems \cite{Rud}, and we
can solve it by the way of {\it Lagrange multipliers}.

Therefore, in the non-entanglement strategy, we have obtained the
minimum error probability for discriminating  two single qubit
quantum operations. On the other hand, in the entanglement
strategy, the minimum error probability $P^{'}_E({\cal E}_1,{\cal
E}_2)$ has not been evaluated. However, our result  implies a
general upper bound for that, by noting that $P_E^{'}({\cal
E}_1,{\cal E}_2)\leq P_E({\cal E}_1,{\cal E}_2)$.

In Eq.~\eqref{Pe}, there are some nontrivial  special cases worthy
of being pointed out:

 (i) When
$|p_1-p_2|\geq\max_{||\Vec{r}||=1}||M\Vec{r}+\Vec{c}||$, we have
that \begin{align} P_E({\cal E}_1,{\cal E}_2)=\min\{p_1, p_2\},
\end{align}
which implies that in this case no  input state and no measurement
are needed, and the minimum error can always be achieved  by just
guessing the most likely operation. We note that similar insight
was also gained for discrimination of quantum states by
\cite{Hunt}, which found that for some set of states, measurement
does not aid minimum-error discrimination, and the minimum error
probability can always be achieved by guessing the most likely
states.

 (ii)
 In particular, when
$\Vec{c}_1=\Vec{c}_2=0$, the minimum error probability can be
evaluated as follows.
 \begin{align}
\begin{split}&P_E({\cal E}_1,{\cal
E}_2)\\=&\frac{1}{2}\Big[1-\max\Big\{|p_1-p_2|,
\max_{||\Vec{r}||=1}||(p_1M_1-p_2M_2)\Vec{r}||\Big\}\Big]\\
=&\frac{1}{2}\Big[1-\max\Big\{|p_1-p_2|,
||p_1M_1-p_2M_2||_2\Big\}\Big],
\end{split}\label{P}
\end{align}
 where $||A||_2$ denotes the {\it spectral norm} \cite{Horn} of matrix
 $A$ as:
 \begin{align}
\begin{split}
||A||_2&=\max\{\sqrt{\lambda}: \lambda\text{ is an eigenvalue of~}  A^\dagger A\}\\
&=\max_{||x||=1}||Ax||.
\end{split}
\end{align}

 In Ref.~\cite{Nie}, some important single qubit quantum operations
 were
 introduced, and they are  bit flip,  phase
flip,  bit-phase flip,  depolarizing, phase damping, and amplitude
damping channels. One can refer to \cite{Nie} for their
operator-sum representations, and below we  visualize them as the
Bloch vector transformations in turn:
\begin{align}\begin{split}
&BF: (r_x,r_y,r_z)\rightarrow(r_x, (2p-1)r_y,(2p-1)r_z),\\
&PF: (r_x,r_y,r_z)\rightarrow ((2p-1)r_x, (2p-1)r_y,r_z),\\
&BPF:(r_x,r_y,r_z)\rightarrow ((2p-1)r_x, r_y,(2p-1)r_z),\\
&DE:(r_x,r_y,r_z)\rightarrow ((1-p)r_x, (1-p)r_y,(1-p)r_z),\\
&PD:(r_x,r_y,r_z)\rightarrow (r_x\sqrt{1-\lambda},
r_y\sqrt{1-\lambda},r_z),\\
&AD:(r_x,r_y,r_z)\rightarrow (r_x\sqrt{1-\lambda},
r_y\sqrt{1-\lambda},r_z(1-\lambda)+\lambda).\end{split}
\end{align}
Therefore, by using Eq.~\eqref{Pe} (or Eq.~\eqref{P}) and the
above transformations, it is easy to get the minimum error
probability for discriminating the above quantum operations.

In fact, bit flip, phase flip, bit-phase flip, and depolarizing
channels are generalized by the more general quantum
operations---Pauli channels defined as:
\begin{align}
{\cal E}(\rho)=\sum_{i=0}^3 q_i\sigma_i\rho\sigma_i,
\end{align}
where
$\{\sigma_0,\sigma_1,\sigma_2,\sigma_3\}=\{I,\sigma_x,\sigma_y,\sigma_z\}$
are the Pauli operators \cite{Nie}, $\sum_iq_i=1$, and $q_i\geq 0$
for any $i$.
 Discrimination of Pauli channels was  discussed in detail for both entanglement strategy and non-entanglement strategy by  \cite{MFS1, MFS2},
 and the minimum error probability $P_E({\cal E}_1,{\cal
 E}_2)$ for them was derived by an elaborate calculation.
 However,   below we will give a more
 intuitional and visual derivation of $P_E({\cal E}_1,{\cal
 E}_2)$, based on  the Bloch representation. At starting, we give
 a lemma as follows.

\noindent\textit{Lemma 2.} Let ${\cal E}(\rho)=\sum_{i=0}^3
q_i\sigma_i\rho\sigma_i$ be the Pauli channel on a single qubit.
Then ${\cal E}$ corresponds to the Bloch vector transformation as
follows:
\begin{align}
\Vec{r}\rightarrow\Vec{r}\hskip 0.5mm'=M\Vec{r}
\end{align}
where $M=diag\{\Delta_1,\Delta_2,\Delta_3\}$, and
$\Delta_i=2(q_0+q_i)-1$.

\textit{ Proof.} Firstly, note that for the Pauli operators
\cite{Nie}, we have
$\sigma_i\sigma_j\sigma_i=\begin{cases}-\sigma_j & j\neq i\\
\sigma_i & j=i\end{cases}$ for $i,j=1,2,3$. Then we get that
\begin{align} \begin{split}{\cal
E}(\rho)=&\sum_{i=0}^3q_i\sigma_i\frac{I+\Vec{r}\cdot\Vec{\sigma}}{2}\sigma_i\\
=&\frac{1}{2}\Big[I+(q_0+q_1-q_2-q_3)r_1\sigma_1\\
&+(q_0-q_1+q_2-q_3)r_2\sigma_2\\
&+(q_0-q_1-q_2+q_3)r_3\sigma_3\Big] \\
=&\frac{1}{2}[I+(M\Vec{r})\cdot\Vec{\sigma}],
\end{split}\end{align} where  $M$ is defined as above. This
completes the proof.\qed

Now, for the two Pauli channels
\begin{align}
{\cal E}_1(\rho)=\sum_{i=0}^3q^{(1)}_i\sigma_i\rho\sigma_i,
~~\text{and}~~ {\cal
E}_2(\rho)=\sum_{i=0}^3q^{(2)}_i\sigma_i\rho\sigma_i,
\end{align}
 by Lemma 2, we know that ${\cal E}_1$ and ${\cal
E}_2$ have 2-tuple representations $(M_1,\Vec{0})$ and
$(M_2,\Vec{0})$, respectively. Then by Eq.~\eqref{P}, we have
 \begin{align}
\begin{split}
&P_E({\cal E}_1,{\cal
E}_2)\\
=&\frac{1}{2}\Big[1-\max\{|p_1-p_2|,
||p_1M_1-p_2M_2||_2\}\Big]\\
=&\frac{1}{2}\Big[1-\max\{|p_1-p_2|, C\}\Big]\label{PPe},
\end{split}
 \end{align}
where \begin{align}\begin{split}
C=\max\Big\{&|r_0+r_1-r_2-r_3|, |r_0+r_2-r_1-r_3|,\\
&|r_0+r_3-r_1-r_2|\Big\},
\end{split}\end{align}
and $r_i=p_1q^{(1)}_i-p_2q^{(2)}_i$ for $i=0,1,2,3$.

In Eq.~\eqref{PPe}, if $C\leq |p_1-p_2|$, then
$P_E=\min\{p_1,p_2\}$, and thus, no exploring input state is
needed as pointed out before. Else,  the optimal exploring input
state has Bloch vector that is  the eigenvector corresponding to
the largest eigenvalue of $p_1M_1-p_2M_2$.

We note that discrimination of Pauli channels was already
discussed in \cite{MFS1, MFS2}, where the minimum error
probability in the non-entanglement strategy was given as
\begin{align}
P_E=\frac{1}{2}(1-M),\label{PPe2}
\end{align}
where
 \begin{align}\begin{split}
M=\max\Big\{&|r_0+r_3|+|r_1+r_2|, |r_0+r_1|+|r_2+r_3|,\\
&|r_0+r_2|+|r_1+r_3|\Big\}.
\end{split}\end{align}
 In fact, one can verify that  Eqs.~\eqref{PPe} and \eqref{PPe2} are equivalent by using Lemma 1 and by noting that $p_1-p_2=r_0+r_1+r_2+r_3$.
 However, as we can see, our method is
based on the Bloch representation of  single qubit systems, and
thus, it is more intuitional and visual than the way used in
\cite{MFS1, MFS2}.
\section{Perfect discrimination of quantum operations} \label{3}
In this section, we consider the condition for perfect
discrimination of two quantum operations. Preliminarily, it is
useful to introduce the notion of {\it numerical range}
\cite{Horn2}. Consider a quantum system with a finite dimensional
state space ${\cal H}$. Let $A$ be a linear operator acting on
${\cal H}$. The {\it numerical range} (or  the {\it field of
values}) \cite{Horn2} of $A$ is defined as
\begin{align}
F(A)=\{\langle\psi|A|\psi\rangle: |\psi\rangle\in{\cal H}
~\text{with}~ \langle\psi|\psi\rangle=1\}.
\end{align}
 The numerical range is an important set that
captures some information about  linear operators. Specially, for
normal operator $A$, the numerical range is the convex hull of the
eigenvalues of $A$, whereas no similar analytical characterization
of numerical range is known for general linear operators.

A nonzero $|\psi\rangle\in {\cal H}$ is said to be an {\it
isotropic vector} for a given linear operator $A$ on ${\cal H}$ if
$\langle\psi|A|\psi\rangle=0$, and $|\psi\rangle$ is further said
to be a {\it common isotropic vector} for linear operators
$A_1,\dots,A_m$ if $\langle\psi|A_i|\psi\rangle=0$ for
$i=1,\dots,m$. Using the above notion, we  give a necessary and
sufficient condition for two quantum operations to be perfectly
distinguishable as follows.

 \noindent\textit{Theorem 1.} Given two quantum operations
\begin{align}{\cal E}_1(\rho)=\sum_{i=1}^{n_1}E_i^{(1)}\rho
{E_i^{(1)}}^\dagger, ~~\text{and}~~ {\cal
E}_2(\rho)=\sum_{j=1}^{n_2}E_{j}^{(2)}\rho
{E_{j}^{(2)}}^\dagger.\end{align} Then:
\begin{enumerate}
\item [(i)]${\cal E}_1$ and ${\cal E}_2$ can be perfectly
discriminated with non-entanglement strategy if and only if
 the set $\{{E_{i}^{(1)}}^\dagger E_j^{(2)}: i=1,\dots,n_1$, $j=1,\dots,n_2\}$
has a common isotropic vector $|\psi\rangle\in{\cal H}$.
 \item[(ii)]${\cal E}_1$ and ${\cal E}_2$ can be perfectly
discriminated with entanglement strategy  if and only if
 the set
$\{{E_{i}^{(1)}}^\dagger E_j^{(2)}\otimes I: i=1,\dots,n_1$,
$j=1,\dots,n_2\}$ has a common isotropic vector
$|\psi\rangle\in{\cal H}\otimes{\cal K}$

\end{enumerate}

 \textit{ Proof.} We have a detailed proof for part (i), and the proof for part (ii) is similar.  We know that the minimum error probability
 can always be achieved by a pure input state. Then, that ${\cal E}_1$ and ${\cal
 E}_2$ can be perfectly discriminated is equivalent to that there
 exists a pure state $|\psi\rangle\in{\cal H}$, such that the output density
 operators are mutual orthogonal, i.e., they have mutual orthogonal
 support. The support of density
 operator $\rho$, denoted by $supp(\rho)$, is defined as the space
 spanned by the eigenvectors corresponding to the no-zero
 eigenvalues of $\rho$. We denote
  \begin{align}\begin{split}
  \rho_1\equiv{\cal
 E}_1(|\psi\rangle\langle\psi|)=\sum_{i=1}^{n_1}E_i^{(1)}|\psi\rangle\langle\psi| {E_i^{(1)}}^\dagger,\\
  \rho_2\equiv{\cal
 E}_2(|\psi\rangle\langle\psi|)=\sum_{j=1}^{n_2}E_{j}^{(2)}|\psi\rangle\langle\psi|
{E_{j}^{(2)}}^\dagger.
 \end{split}\end{align}
 Then it follows that
\begin{align}\begin{split}
supp(\rho_1)=span\{E_i^{(1)}|\psi\rangle: i=1,\dots, n_1\},\\
supp(\rho_2)=span\{E_j^{(2)}|\psi\rangle: j=1,\dots, n_2\}.
\end{split}\end{align}
Here it is readily seen that  $supp(\rho_1)\perp supp(\rho_2)$ if
and only if $E_i^{(1)}|\psi\rangle\perp E_j^{(2)}|\psi\rangle$,
i.e., $\langle\psi|{E_{i}^{(1)}}^\dagger E_j^{(2)}|\psi\rangle=0$
for any $i$ and $j$. Therefore, we have proven the first part of
this theorem. For the strategy with entangled input, we have a
similar proof, and thus we omit that. \qed

In Theorem 1, it is easy to see that two quantum operations that
can be perfectly discriminated by non-entanglement strategy are
necessary to be perfectly distinguishable by entanglement
strategy. However, the contrary implication is not true by noting
the later example.

 Notably, when two unitary operations $U_1$ and $U_2$
are considered,  the condition for perfect discrimination between
them reduces  to that $\langle\psi|U_1^\dagger U_2|\psi\rangle=0$
for some input state $|\psi\rangle\in{\cal H}$, i.e., the polygon
of the eigenvalues of $U_1^\dagger U_2$ encircles the origin,
which was discussed in Refs. ~\cite{Acin,Paris}.

Theorem 1 implies that it is generally difficult to  decide
whether two quantum operations are perfectly distinguishable.
However, it will be easy in some special cases as we will show
below.   In the following, we consider the condition for two Pauli
channels to be perfectly distinguishable. For that, we discuss a
more general case, that is, discriminating the following quantum
operations \cite{MFS1,MFS2}:
\begin{align}
{\cal E}_i(\rho)=\sum_{n=0}^{d^2-1} q^{(i)}_nU_n\rho U_n^\dagger,
~~\sum_nq^{(i)}_n=1,~ \text{and}~ q^{(i)}_n\geq 0,\label{GP}
\end{align}
with $\text{Tr}(U_m^\dagger U_n)=d\delta_{mn}$. From the above
form, we know that ${\cal E}_i$ has operator-sum element set
$\{\sqrt{q^{(i)}_n} U_n\}$.  Also, we can see that when the
orthogonal set $\{U_n\}$ is fixed,  ${\cal E}_i$ is unique
determined by the unit $d^2$-dimensional vector ${\bf
q^{(i)}}=(\sqrt{q^{(i)}_0},\dots,\sqrt{q^{(i)}_{d^2-1}})$, which
is called the {\it characteristic vector} of ${\cal E}_i$. As we
can see, when $d=2$, these operations reduce to the Pauli channels
\cite{MFS1,MFS2}. Thus, we call these quantum operations defined
above as {\it general Pauli channels} (GPCs). For these general
Pauli channels, we have the following theorem.

\noindent\textit{Theorem 2.} Two GPCs  ${\cal E}_1$ and ${\cal
E}_2$ are perfectly distinguishable with entangled strategy, if
and only if their characteristic vectors are orthogonal.

\textit{ Proof.} Suppose that ${\cal E}_1$ and ${\cal E}_2$ have
characteristic vectors ${\bf q^{(1)}}=[\sqrt{q^{(1)}_i}]$ and
${\bf q^{(2)}}=[\sqrt{q^{(2)}_i}]$, respectively. Then from
Theorem 1, we know that ${\cal E}_1$ and ${\cal E}_2$ are
perfectly distinguishable with entangled strategy if and only if
the following hold:
\begin{align}
\sqrt{q^{(1)}_iq^{(2)}_j}\langle\psi|U^\dagger_iU_j\otimes
I|\psi\rangle=0\label{Con1}
\end{align}
for $i,j=0,\dots, d^2-1$ and some $|\psi\rangle\in {\cal
H}\otimes{\cal K}$. Let $i=j$ in Eq.~\eqref{Con1}. Then  we have
 \begin{align}\sqrt{q^{(1)}_iq^{(2)}_i}=0, ~~\text{for}~~ i=0,\dots,d^2-1,\label{Con}\end{align}
 which is equivalent to ${\bf q^{(1)}}\perp{\bf q^{(2)}}$. Now we
 have verified the necessity.

 Next  suppose that  ${\bf q^{(1)}}\perp{\bf
 q^{(2)}}$ holds, i.e., Eq.~\eqref{Con} holds. Let us recall the notation of Ref.~\cite{GMD} for bipartite state
\begin{align}
|A\rangle\rangle=\sum_{n,m}A_{nm}|n\rangle\otimes|m\rangle,
\end{align} and there are properties:\\
(i) $A\otimes B|C\rangle\rangle=|ACB^T\rangle\rangle$;\\
 (ii)
$\langle\langle A|B\rangle\rangle=\text{Tr}(A^\dagger B)$;
\\(iii) any maximal entangled state can be represented as
$\frac{1}{\sqrt{d}}|V\rangle\rangle$, where $V$ is a unitary
matrix.

 Then using the notation introduced above,  for any maximal entangled state
 $|\psi\rangle=\frac{1}{\sqrt{d}}|V\rangle\rangle$, and for any $i,j$, we have
 \begin{align}\begin{split}
&\sqrt{q_i^{(1)}q_j^{(2)}}\langle\psi|U_i^\dagger U_j\otimes
I|\psi\rangle\\
=&\frac{1}{d}\sqrt{q_i^{(1)}q_j^{(2)}}\langle\langle
U_iV|U_jV\rangle\rangle\\
=&\frac{1}{d}\sqrt{q_i^{(1)}q_j^{(2)}}\text{Tr}(V^\dagger
U_i^\dagger U_j
V)\\
=&\sqrt{q_i^{(1)}q_j^{(2)}}\delta_{ij}\\
=&0\end{split}
 \end{align}
Thus, from Theorem 1, ${\cal E}_1$ and ${\cal E}_2$ are perfectly
distinguishable by the maximal entangled state $|\psi\rangle$.
This completes the proof.\qed

Notably, the condition given in the above theorem is also a
necessary condition for ${\cal E}_1$ and ${\cal E}_2$ to be
perfectly distinguishable with non-entanglement strategy. However,
it is not sufficient. An example of this case is two channels of
the form
\begin{align}
{\cal
E}_1(\rho)=\sum_{\alpha\neq\beta}q_\alpha\sigma_\alpha\rho\sigma_\alpha,~{\cal
E}_2(\rho)=\sigma_\beta\rho\sigma_\beta,
\end{align}
with $q_\alpha\neq0$, as {\it a priori} probability.
\section{Conclusions}\label{4}
In this work, we addressed the problem of discrimination between
quantum operations. For the single qubit quantum operations, we
obtained the exact minimum error probability with non-entanglement
strategy. For the Pauli channels discussed in \cite{MFS1, MFS2},
we gave a more intuitional and visual solution to their
discrimination problem. We gave a necessary and sufficient
condition for two quantum operations to be perfectly
distinguishable, and as an application, we found that two
generalized Pauli operations are perfectly distinguishable if and
only if their characteristic vectors are orthogonal.

This work is supported  by the National Natural Science Foundation
(Nos. 90303024, 60573006), the Research Foundation for the
Doctorial Program of Higher School of Ministry of Education (No.
20050558015), and the Natural Science Foundation of Guangdong
Province (No. 031541) of China.


\begin{thebibliography}{ABCD}
\bibitem {Hel} C. W. Helstrom, {\it Quantum Detection and Estimation
Theory}
(Academic Press, New York, 1976).
\bibitem {Four}I. D. Ivanovic, Phys. Lett. A {\bf 123}, 257 (1987); D. Dieks, {\it ibid}.
{\bf 126}, 303 (1988); A. Peres, {\it ibid}. {\bf 128}, 19 (1988);
G. Jaeger and A. Shimony, {\it ibid}. {\bf 197}, 83 (1995); A.
Chefles, {\it ibid}. {\bf 239}, 339 (1998).

\bibitem{Bergou04}  J. Bergou, U. Herzog, and M. Hillery,
 {\it Quantum State Estimation}, Lecture Notes in Physics Vol. 649
(Springer, Berlin, 2004), p. 417; A. Chefles, {\it ibid}. p. 467.
\bibitem {Nie} M. A. Nielsen, and I. L. Chuang, {\it Quantum Computation and Quantum
Information} (Cambridge University Press, Cambridge, 2000)

\bibitem {Childs} A. M. Childs, J. Preskill, and J. Renes, J. Mod. Opt. {\bf 47}, 155
(2000).
\bibitem {Acin} A. Ac\'{i}n, Phys. Rev. Lett. {\bf 87}, 177901 (2001).
\bibitem {Paris} G. M. D'Ariano, P. Lo Presti, and M. G. A. Paris,
Phys. Rev. Lett. {\bf  87}, 270404 (2001).
\bibitem {Duan}R. Y. Duan, Y. Feng, and M. S. Ying,  Phys. Rev. Lett. {\bf 98}, 100503 (2007).
\bibitem{MFS1} M. F. Sacchi, J. Opt. Soc. Am. B {\bf 7}, S333 (2005).
\bibitem {MFS2} M. F. Sacchi,  Phys. Rev. A {\bf 71}, 062340 (2005).
\bibitem {MFS3} G. M. D'Ariano, M. F. Sacchi, Phys. Rev. A {\bf 72}, 052302 (2005).
\bibitem{Wang} G. Wang, M. S. Ying,  Phys. Rev. A {\bf 73}, 042301 (2006).
\bibitem{Chef07}  A. Chefles, A. Kitagawa, M. Takeoka, M. Sasaki and J.
Twamley, quant-ph/0702245v3.
\bibitem{Rud} W. Rudin, {\it Principles of Mathematical
Analysis} (McGraw-Hill, New York, 1976).
\bibitem{Hunt} K. Hunter, Phys. Rev. A {\bf 68}, 012306 (2003).
\bibitem{Horn} R. A. Horn, C. R. Johnson, {\it Matrix Analysis}
(Cambridge University Press, Cambridge, 1986).
\bibitem{Horn2} R. A. Horn, C. R. Johnson, {\it Topics in Matrix Analysis},
(Cambridge University Press, Cambridge, 1991).
\bibitem{GMD} G. M. D'Ariano, P. Lo Presti, and M. F. Sacchi, Phys. Lett. A {\bf 272}, 32 (2000).

\end{thebibliography}
\end{document}